\documentclass[conference]{IEEEtran}
\IEEEoverridecommandlockouts
\usepackage{cite}
\usepackage{amsmath,amssymb,amsfonts}
\usepackage{algorithmic}
\usepackage{graphicx}
\usepackage{textcomp}
\usepackage{xcolor}
\usepackage{url}
\usepackage{array}
\usepackage{booktabs}
\usepackage{listings}
\usepackage{hyperref}
\usepackage{seqsplit}
\usepackage{orcidlink}
\usepackage{courier} 

\def\BibTeX{{\rm B\kern-.05em{\sc i\kern-.025em b}\kern-.08em
    T\kern-.1667em\lower.7ex\hbox{E}\kern-.125emX}}

\begin{document}

\title{An Organization-Scoped LLM Agent Runtime Architecture for Regulated Cybersecurity Operations\\
\thanks{This work was supported by the European Union under the DIGITAL EUROPE Programme, grant agreement No.~101249596 (CyberAId).}
}

\author{
\begin{tabular}{ccc}
\begin{minipage}[t]{0.30\textwidth}\centering
\textbf{1\textsuperscript{st} George Fatouros\,\orcidlink{0000-0001-6843-089X}} \\
\textit{Innov-Acts Ltd} \\
Nicosia, Cyprus \\
\href{mailto:gfatouros@innov-acts.com}{gfatouros@innov-acts.com}
\end{minipage} &
\begin{minipage}[t]{0.30\textwidth}\centering
\textbf{2\textsuperscript{nd} Georgios Makridis} \\
\textit{Dept. of Digital Systems} \\
\textit{University of Piraeus}\\
Athens, Greece \\
\href{mailto:gmakridis@unipi.gr}{gmakridis@unipi.gr}
\end{minipage} &
\begin{minipage}[t]{0.35\textwidth}\centering
\textbf{3\textsuperscript{rd} George Kousiouris} \\
\textit{Dept. of Informatics and Telematics} \\
\textit{Harokopio University}\\
Athens, Greece \\
\href{mailto:gkousiou@hua.gr}{gkousiou@hua.gr}
\end{minipage} \\[6em]
\multicolumn{3}{c}{
\begin{tabular}{cc}
\begin{minipage}[t]{0.30\textwidth}\centering
\textbf{4\textsuperscript{th} John Soldatos} \\
\textit{Innov-Acts Ltd}\\
Nicosia, Cyprus \\
\href{mailto:jsoldatos@innov-acts.com}{jsoldatos@innov-acts.com}
\end{minipage} &
\begin{minipage}[t]{0.30\textwidth}\centering
\textbf{5\textsuperscript{th} Dimosthenis Kyriazis} \\
\textit{Dept. of Digital Systems} \\
\textit{University of Piraeus}\\
Athens, Greece \\
\href{mailto:dimos@unipi.gr}{dimos@unipi.gr}
\end{minipage}
\end{tabular}
}
\end{tabular}
}

\maketitle

\begin{abstract}
Regulated cybersecurity workflows lack a runtime substrate that enforces organization-level scope across retrieval, tool calls, memory, findings, reports, and audit while remaining model-agnostic and locally deployable. Recent large language model (LLM) agent systems report strong results on isolated cybersecurity tasks, yet they do not by themselves define an auditable platform architecture for regulated security operations centre (SOC) and compliance workflows, where a single analyst may trigger actions that bind the organization, and where the runtime must integrate with existing SIEM/XDR stacks as a primary source of context and alert-driven triggers rather than operate as a standalone analytical layer. This paper proposes an organization-scoped LLM agent runtime architecture for financial cybersecurity. The contribution is a typed Security Context that is created at every entry point---including SIEM/XDR notifications ingested as first-class triggers---and enforced at every component boundary, combined with a shared Runtime Core, logical specialist subagents, a governed Tool Adapter Layer exposing SIEM/XDR query, enrichment, and response primitives under uniform policy and audit, structured findings with evidence references, tiered human-in-the-loop (HITL) gates, and append-only audit. Model Context Protocol (MCP), extended telemetry, digital twins for pentesting, graph retrieval, and federated knowledge sharing are treated as optional extension paths rather than mandatory runtime assumptions. We describe an implementable slice as the architecture's testability surface, and we propose a falsifiable evaluation plan with metric-level pass criteria for architecture readiness, security-policy enforcement, evidence traceability, output quality, and operational observability.
\end{abstract}


\section{Introduction}\label{sec:intro}

Financial security operations centres (SOCs) operate under a mismatch between the amount of available telemetry and the amount of reasoning that can be applied to it. Security information and event management (SIEM), extended detection and response (XDR), vulnerability scanners, ticketing systems, regulatory evidence stores, and cyber threat intelligence (CTI) feeds already produce useful signals, yet many incidents still require analysts to connect events across systems, assets, identities, policies, and regulatory obligations under deadlines defined by the DORA,\footnote{\href{https://eur-lex.europa.eu/eli/reg/2022/2554/oj}{Digital Operational Resilience Act Regulation (EU) 2022/2554}.} NIS2,\footnote{\href{https://eur-lex.europa.eu/eli/dir/2022/2555/oj}{Network and Information Security Directive 2 (EU) 2022/2555}.} GDPR,\footnote{\href{https://eur-lex.europa.eu/eli/reg/2016/679/oj}{General Data Protection Regulation (EU) 2016/679}.} PSD2,\footnote{\href{https://eur-lex.europa.eu/eli/dir/2015/2366/oj}{Payment Services Directive 2 (EU) 2015/2366}.} and AI Act.\footnote{\href{https://eur-lex.europa.eu/eli/reg/2024/1689/oj}{EU Artificial Intelligence Act (EU) 2024/1689}.}

LLM agents are now practical building blocks for tool use, retrieval, multi-agent coordination, and cybersecurity analysis~\cite{wang2024survey,fatouros2025marketsense, ref-survey-agentic-sec}. However, narrow demonstrations such as vulnerability exploitation, intrusion classification, phishing analysis, or isolated incident-response support do not define a regulated platform architecture. A financial SOC runtime must persist state, enforce authorization, scope retrieval, call tools through deterministic contracts, assemble evidence references, support human-in-the-loop (HITL) review, and produce audit trails that can be inspected after the fact \cite{fatouros2026cyberaid}.

Contemporary agent platforms separate model providers from the runtime, expose tools through schemas, delegate to subagents, and verify outputs through plans and reviews~\cite{ref-codex-loop,ref-claude-code,ref-cursor-best-practices,ref-openclaw-runtime}, but assume a personal-copilot deployment. Regulated cybersecurity needs the same separation of concerns under \emph{organization-level} authority: the user may be one analyst, but the agent accesses sensitive evidence, regulated systems, and high-impact workflows.

This paper proposes an organization-scoped LLM agent runtime architecture. The central design choice is a typed \emph{Security Context} created at every entry point and enforced uniformly at agent invocation, retrieval, graph queries, tool calls, memory access, finding/report generation, and UI visibility. Around it we organize a Runtime Core, governed Tool Adapter Layer, Context Broker, logical specialist subagents, structured findings with evidence references, tiered HITL gates, and append-only audit. We describe the architecture as an implementable slice and propose a falsifiable evaluation plan, without claiming production validation.

The paper makes four contributions: (i) an organization-scoped agent-runtime architecture with an implementable slice that exercises every runtime contract without the full CyberAId platform stack; (ii) a typed Security Context with documented enforcement points across retrieval, graph, tool, memory, finding, report, and UI surfaces; (iii) a governance model combining the Security Context, scoped retrieval, typed Tool Adapter Layer, tiered HITL, structured findings with evidence references, and append-only audit; and (iv) a falsifiable evaluation plan with metric-level pass criteria covering architecture readiness, policy enforcement, evidence traceability, output quality, and operational observability.

\begin{figure}[t]
\centering
\includegraphics[width=\columnwidth]{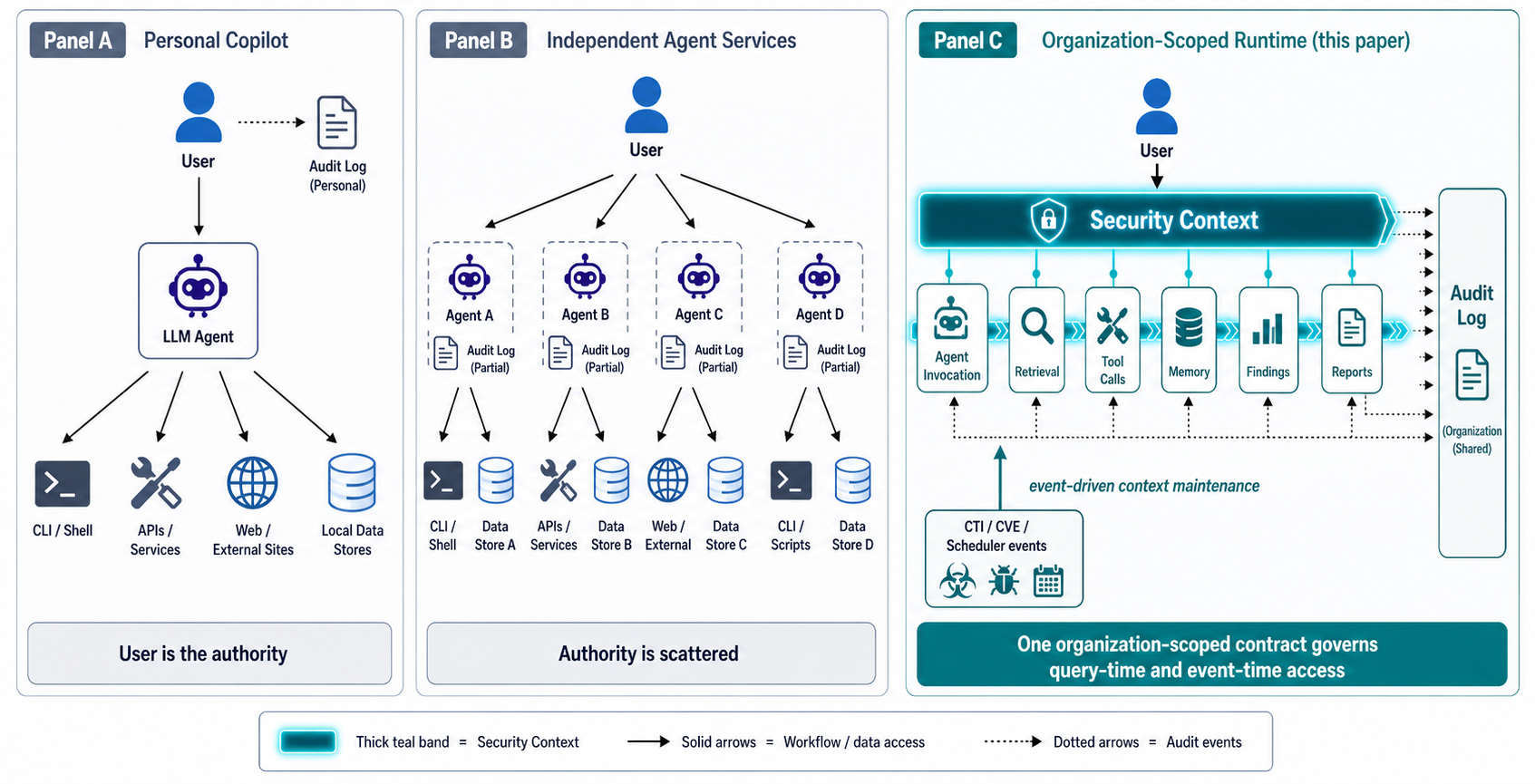}
\caption{Three deployment patterns for LLM agents acting on behalf of a user. Personal copilots (A) and independent agent services (B) leave organization-level authority, audit, and event-time context maintenance to the surrounding deployment. The organization-scoped runtime proposed in this paper (C) makes a typed Security Context the unifying contract across query-time access, event-driven updates, and audit.}
\label{fig:concept}
\end{figure}

\noindent\textbf{Novelty:} Unlike generic multi-agent orchestration frameworks~\cite{ref-autogen} and personal-copilot agent platforms~\cite{ref-codex-loop,ref-claude-code}, the runtime proposed here treats organization scope as a first-class architectural primitive: the Security Context is the sole object whose presence is required for retrieval, graph queries, tool calls, memory access, finding generation, report generation, and UI visibility, and its absence is a deny condition rather than a default. The same contract governs event-driven context updates and analyst-initiated queries, making the runtime a self-updating operational substrate rather than a passive retrieval layer. The single-user-but-organization-scoped framing has not been formalized as a runtime contract for regulated cybersecurity operations. Figure~\ref{fig:concept} summarises the framing visually.

\section{Background and Related Work}\label{sec:related}

\textit{LLM-agent runtimes.} Two patterns dominate practical agent construction: ReAct-style tool calling~\cite{ref-react} and plan-and-execute~\cite{wang2024survey}, composing into multi-agent orchestrations~\cite{ref-autogen} over RAG~\cite{lewis2020retrieval}, chain-of-thought~\cite{wei2022chain, fatouros2025can}, and programmatic tool use~\cite{schick2023toolformer}. Coding-agent platforms formalize agent loops, schema-exposed tools, subagents, project rules, and review gates~\cite{ref-claude-code,ref-cursor-best-practices,ref-openclaw-runtime}, but target individual developers and leave authority, audit, and tenancy to the deployment.

\textit{Cybersecurity LLM-agent systems.} Frontier models autonomously exploit one-day vulnerabilities~\cite{ref-llm-exploits}; multi-agent attack teams improve over single-agent baselines on zero-day exploitation~\cite{zhu2026teams}; LLM-assisted SAST raises vulnerability-detection F1~\cite{ref-iris}; intrusion-detection agents reach high F1~\cite{ref-idsagent}. Hybrid systems grounding LLM reasoning in classical detection --- SIEM-grounded triage~\cite{ref-cortex}, agentic-RAG classification~\cite{blefari2025cyberrag}, graph-augmented threat intelligence~\cite{ref-ctikg}, and knowledge-graph phishing detection~\cite{ref-knowphish} --- consistently outperform end-to-end pipelines~\cite{ref-survey-agentic-sec,ref-he-survey}.

\textit{Agent-protocol and tool-use security:} The agentic threat surface includes prompt injection, knowledge-base poisoning, tool-call hijacking, and protocol exploits~\cite{ferrag2025prompt,ref-he-survey}; knowledge-base poisoning reaches high success against unprotected retrieval~\cite{ref-survey-llm-sec}; MCP security guidance highlights authorization, consent, confused-deputy, and tool-description injection~\cite{ref-mcp-security}.

\textit{Regulatory frameworks:} DORA, NIS2, GDPR, PSD2, and the AI Act define obligations on incident reporting, evidence preservation, oversight, and resilience for financial institutions. Meanwhile, research works on LLM agents in financial cybersecurity remains sparse~\cite{wu2024threatmodeling}.

\subsection{Design Requirements}\label{sec:requirements}

The architecture is guided by five requirements that differ from those of a personal assistant or generic multi-agent framework. \textit{Local and model-agnostic deployment:} the Runtime Core is local-first and the Model Gateway abstracts local and private API endpoints, so agent logic does not depend on a specific provider; external context sources (CTI feeds, regulatory portals) operate under the institution's audited egress policy. \textit{Organization-scoped operation:} the interface may serve one analyst, but the system acts under organization-level authority and must represent roles, delegated permissions, data classifications, evidence boundaries, escalation rules, regulatory obligations, and durable institutional memory --- not via prompts alone. \textit{Governed context and tool access:} retrieval and tool use are the main trust boundaries; every access must be mediated by a Security Context, and tool calls must use deterministic schemas, least-privilege permissions, validation, and audit logging~\cite{lewis2020retrieval,schick2023toolformer,ref-mcp-security}. \textit{Auditability and HITL:} findings and reports must carry evidence, retrieval, tool-call, and model references with confidence and HITL status; high-impact, low-confidence, or irreversible actions route to human review. \textit{Maintainability and extension:} the Runtime Core remains stable while capabilities are added through connector, subagent, skill, and context packs.

\section{Architecture and Organization-Scoped Agent Model}\label{sec:architecture}

\begin{figure}[ht]
\centering
\includegraphics[width=\columnwidth]{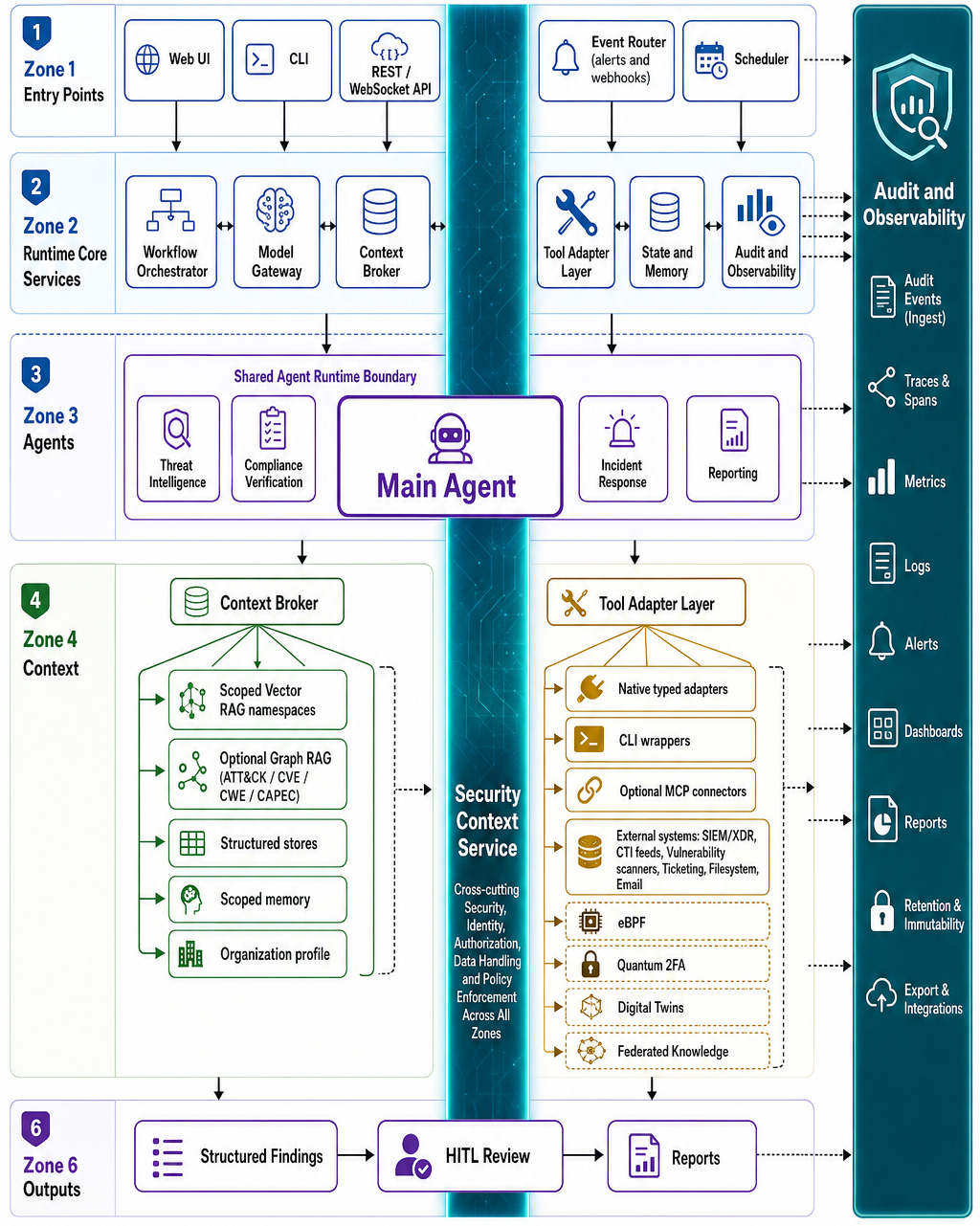}
\caption{Organization-scoped LLM agent runtime architecture. Entry points share a Runtime Core that owns model access, context access, tool mediation, state, memory, audit, and observability. Specialist subagents are logical actors inside the runtime, not independently deployed services by default. The Security Context (Section~\ref{sec:secctx}, Table~\ref{tab:enforcement}) is the cross-cutting enforcement object, and the Audit and Observability lane on the right receives events from every Runtime Core service, every adapter, every retrieval call, and every output stage.}
\label{fig:runtime-architecture}
\end{figure}

Figure~\ref{fig:runtime-architecture} shows the proposed runtime. The architecture uses one shared Runtime Core and multiple logical specialist subagents. This avoids the complexity of deploying each agent as an independent service while preserving specialization and independent invocation.

\textit{Runtime Core.} The stable platform boundary; it owns the API/UI backend, CLI surface, event router, scheduler, workflow orchestrator, Security Context service, Model Gateway, Context Broker, Tool Adapter Layer, state, memory, audit, and observability. Subagents own role instructions and logical namespaces but never storage, policy enforcement, lifecycle, or audit.

\textit{Main Agent and specialist subagents.} The Main Agent is the analyst-facing coordinator: it answers directly, dispatches one specialist, runs parallel workflows, sequences findings, and routes actions to HITL review. Specialist subagents are logical actors with role instructions, allowed tools, RAG namespaces, optional graph scopes, output schemas, and confidence/HITL thresholds~\cite{ref-autogen}. An implementable slice starts with Threat Intelligence, Compliance Verification, Incident Response, and Reporting.

\textit{Model Gateway and Context Broker.} The Model Gateway routes inference to local or private endpoints and tracks model versions; the Context Broker mediates document RAG, optional graph RAG, structured stores, scoped memory, and live context. Model selection is independent from context governance.

\textit{Tool Adapter Layer.} Three peer adapter types (native typed adapters, CLI wrappers, MCP connectors) all subject to the same Security Context enforcement, schema validation, provenance checks, and audit logging. Native and CLI adapters are operationally preferred for high-audit paths; MCP suits dynamic-discovery paths.

\textit{State, memory, and observability.} State covers workflow checkpoints, pending approvals, retries, deadlines, cancellations, and resumability; memory covers scoped agent memory and approved institutional memory; observability covers model latency, token use, tool success, retrieval coverage, confidence distributions, finding acceptance/rejection, HITL rates, connector health, scheduler backlog, and drift.

\subsection{Event-driven context maintenance}\label{sec:event-driven}
The runtime is not only a query-time retrieval layer: alerts, webhooks, scheduled routines, connector-health events, CTI/CVE feed changes, and accepted analyst corrections initiate governed workflows that refresh scoped context, graph knowledge, findings, tickets, and approved memory under the same Security Context, schema validation, audit, and HITL tiers as analyst-initiated workflows (Table~\ref{tab:event-driven}). The Event Router and Scheduler are the temporal entry points; every event acquires a Security Context at admission.

\begin{table}[t]
\centering
\caption{Event-driven update paths. Every path inherits the Security Context of the triggering event and is recorded in the audit log.}
\label{tab:event-driven}
\small
\renewcommand{\arraystretch}{1.15}
\newcommand{\auditcode}[2]{\texttt{\footnotesize #1.\allowbreak #2}}
\begin{tabular}{@{}>{\raggedright\arraybackslash}p{0.26\columnwidth}
                  >{\raggedright\arraybackslash}p{0.30\columnwidth}
                  >{\raggedright\arraybackslash}p{0.13\columnwidth}
                  >{\raggedright\arraybackslash}p{0.21\columnwidth}@{}}
\toprule
\textbf{Trigger} & \textbf{Update class} & \textbf{HITL tier} & \textbf{Audit event} \\
\midrule
CTI / CVE feed change                   & Scoped vector and graph index refresh                     & Tier 1            & \auditcode{context}{refresh} \\
Connector-health event                  & Connector-state record; degraded-state alert              & Tier 1            & \auditcode{connector}{health} \\
Scheduled routine (e.g.\ daily posture) & Findings draft; report draft                              & Tier 1 / 2        & \auditcode{routine}{run} \\
Incoming alert / webhook                & Workflow start; finding emission                          & Tier per finding  & \auditcode{workflow}{start} \\
Accepted analyst correction             & Scoped agent-memory write; institutional-memory candidate & Tier 2            & \auditcode{memory}{write} \\
Pack upgrade with new claims            & Manifest re-verification                                  & Tier 3            & \auditcode{pack}{upgrade} \\
\bottomrule
\end{tabular}
\end{table}

No event-driven update bypasses the Security Context, Finding Schema, HITL tier mapping, or audit log; event-time and query-time access are duals of one governance contract. Writes that affect downstream reasoning (e.g.\ institutional-memory updates from accepted findings) require at least Tier~2 review, so paths targeted by knowledge-base poisoning or evidence-injection~\cite{ferrag2025prompt,ref-survey-llm-sec} cannot execute autonomously.

\subsection{Authority and scope}\label{sec:org-model}

A single-user interface does not imply a personal assistant: in regulated cybersecurity, one analyst may trigger workflows that access organization-owned evidence and affect organizational reporting obligations. The runtime must therefore distinguish the human interface from the authority and scope under which the system operates. Table~\ref{tab:personal-vs-org} contrasts a personal coding-agent deployment with the organization-scoped runtime.

\begin{table}[t]
\centering
\caption{Personal coding-agent deployment versus organization-scoped cybersecurity runtime.}
\label{tab:personal-vs-org}
\small
\begin{tabular}{p{0.18\columnwidth} p{0.30\columnwidth} p{0.38\columnwidth}}
\toprule
\textbf{Concern} & \textbf{Personal coding agent} & \textbf{Organization-scoped cyber agent} \\
\midrule
Rules        & Project conventions     & Regulatory, operational, evidence-handling \\
Surface      & Local files, shell, per-workspace memory & SIEM/XDR, scanners, ticketing, identity, compliance evidence; scoped + institutional memory \\
Review       & Correctness, tests      & Correctness, authorization, traceability, regulatory impact, audit \\
Failure mode & Bad commit              & Regulatory finding, evidence loss, customer impact \\
\bottomrule
\end{tabular}
\end{table}

\subsection{Security Context as a runtime contract}\label{sec:secctx}
\begin{figure*}[t]
\centering
\includegraphics[width=\linewidth]{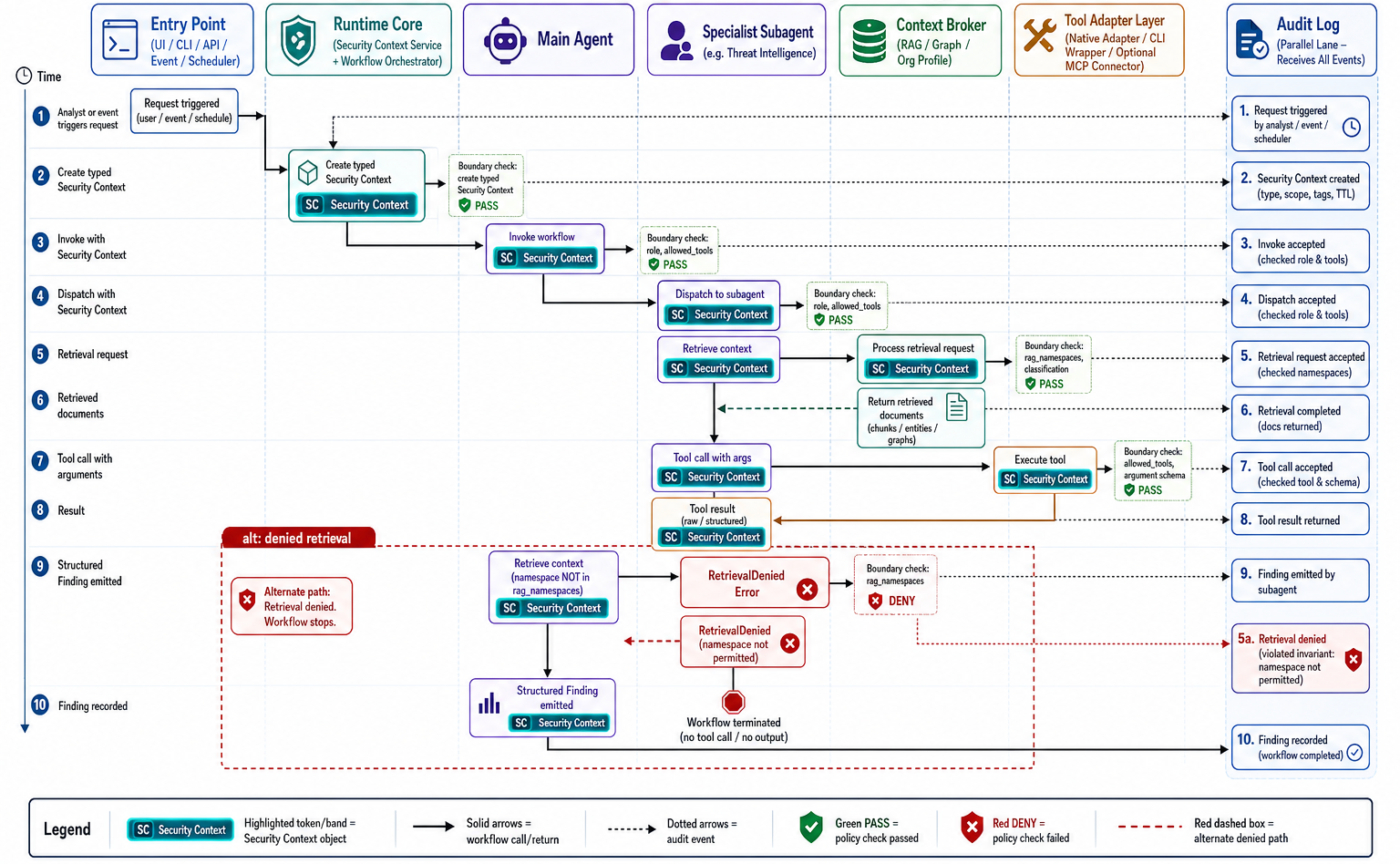}
\caption{Security Context propagation through one workflow. Each boundary check (Table~\ref{tab:enforcement}) emits an audit event; the dashed branch shows a denied retrieval producing \texttt{RetrievalDenied} with the violated invariant logged.}
\label{fig:secctx-propagation}
\end{figure*}

The Security Context is a typed object created at every entry point. Listing~\ref{lst:secctx} gives its schema; the same object is then required at every component boundary listed in Table~\ref{tab:enforcement}, and its absence is a deny condition. Figure~\ref{fig:secctx-propagation} shows how a single Security Context object propagates through a canonical workflow.

\begin{lstlisting}[
  float,
  language={},
  caption={Security Context schema. The runtime treats the object as immutable for the lifetime of a workflow; downstream components may receive narrower views but never broader ones.},
  label={lst:secctx}
]
SecurityContext := {
  org_id          : OrganizationId,
  user_id         : UserId,
  role            : Role,
  data_scope      : Set<DataScope>,
  classification  : ClassificationLevel,
  allowed_tools   : Set<ToolId>,
  rag_namespaces  : Set<NamespaceId>,
  graph_scopes    : Set<GraphScopeId>,
  workflow_origin : Origin, // chat | alert | webhook | schedule | cli | api
  audit_token     : AuditToken,
  created_at      : Timestamp,
  expires_at      : Timestamp
}
\end{lstlisting}

\begin{table}[t]
\centering
\caption{Security Context enforcement points and deny conditions.}
\label{tab:enforcement}
\small
\begin{tabular}{p{0.32\columnwidth} p{0.30\columnwidth} p{0.26\columnwidth}}
\toprule
\textbf{Enforcement point} & \textbf{Invariant checked} & \textbf{Deny error} \\
\midrule
Agent invocation     & \texttt{role}, \texttt{allowed\_tools}                & \texttt{AgentDenied} \\
Document retrieval   & \texttt{rag\_namespaces}, \texttt{classification}     & \texttt{RetrievalDenied} \\
Graph query          & \texttt{graph\_scopes}                                & \texttt{GraphDenied} \\
Tool call            & \texttt{allowed\_tools}, argument schema              & \texttt{ToolDenied} \\
Memory read/write    & \texttt{org\_id}, \texttt{data\_scope}                & \texttt{MemoryDenied} \\
Finding generation   & \texttt{classification}, evidence-reference policy    & \texttt{FindingDenied} \\
Report generation    & finding-set scope, regulatory template scope          & \texttt{ReportDenied} \\
UI visibility        & \texttt{role}, \texttt{data\_scope}                   & \texttt{UIRedacted} \\
\bottomrule
\end{tabular}
\end{table}

The Runtime Core enforces the invariants in Table~\ref{tab:enforcement}; subagents do not implement their own enforcement. Each deny event is recorded in the append-only audit log together with the violated invariant, allowing post-hoc inspection of every denied access.

\noindent\textit{Threat-model scope.} The Security Context defends against scope-violation attacks (cross-org or cross-tenant retrieval, queries outside \texttt{rag\_namespaces} or \texttt{graph\_scopes}, tool calls outside \texttt{allowed\_tools} or with arguments failing schema validation, memory writes outside \texttt{org\_id}/\texttt{data\_scope}, and unaudited execution at any boundary), and pack governance (Section~\ref{sec:pack-governance}) extends the same contract to extension code. Within-scope attacks (prompt injection over authorized retrieval, persuasive but false analyses, tool calls with valid arguments acting on attacker-influenced inputs) are addressed by Runtime Core guardrails layered on top: input/output validation and prompt-injection detection at the Context Broker and Tool Adapter Layer, deterministic schemas, confidence thresholding and source-attribution on findings, and tier-bound HITL review (Table~\ref{tab:hitl}). The append-only audit log is the residual defence required by DORA Article~17 and AI Act Article~14~\cite{ferrag2025prompt,ref-survey-llm-sec,ref-he-survey}.

An organization profile (e.g.\ \texttt{CYBERAID.md}) complements the enforcement layer: a reviewable file containing regulatory profile, asset descriptions, escalation contacts, report preferences, terminology, classification rules, and playbook references. The profile is loaded into prompts and retrieval context but never grants permissions; permissions live only in the Security Context. The model scales from a one-operator deployment to institutional use without rework: additional users acquire their own Security Contexts under the same \texttt{org\_id}, additional roles narrow rather than widen scope, and multi-tenant operation requires adding an explicit \texttt{tenant\_id} field.

\section{Implementation, Findings, HITL, and Audit}\label{sec:slice}

\subsection{Implementable Slice}

This section describes the implementable slice as the architecture's testability surface. The slice below is the testability surface against which the evaluation plan in Section~\ref{sec:evaluation} can be executed.

The architecture can be exercised through a deliberately small slice: a local Runtime Core process; a Model Gateway with one local endpoint and one private API endpoint; a vector store with namespaces and an optional graph interface limited to ATT\&CK/CVE queries; filesystem, webhook, and mock SIEM/alert adapters; the Main Agent plus Threat Intelligence, Compliance Verification, Incident Response, and Reporting subagents; the structured Finding Schema; an append-only audit log over prompts, retrievals, tool calls, findings, approvals, and reports; and one scheduled routine such as a daily posture summary. 

\subsection{Structured findings}\label{sec:findings-hitl}\label{sec:findings}
Specialist subagents communicate with the Main Agent, the Reporting subagent, the dashboards, and the audit log through a single Finding object. The schema is designed to make every claim attributable. Required fields are listed in Table~\ref{tab:finding-schema}; the runtime rejects findings that fail schema validation or whose Security Context cannot be re-resolved.

\begin{table}[t]
\centering
\caption{Required fields of the structured Finding Schema.}
\label{tab:finding-schema}
\small
\renewcommand{\arraystretch}{1.15}
\newcommand{\fld}[1]{\texttt{\seqsplit{#1}}}
\begin{tabular}{@{}>{\raggedright\arraybackslash}p{0.30\columnwidth}
                  >{\raggedright\arraybackslash}p{0.60\columnwidth}@{}}
\toprule
\textbf{Field} & \textbf{Purpose} \\
\midrule
\fld{finding\_id}            & Unique identifier within the runtime. \\
\fld{workflow\_id}           & Owning workflow. \\
\fld{agent\_id}              & Producing subagent. \\
\fld{skill\_ids}             & Skill packs active when the finding was emitted. \\
\fld{timestamp}             & Emission time. \\
\fld{security\_context\_ref}  & Reference to the Security Context under which the finding was produced. \\
\fld{origin}                & Entry-point class (chat, alert, webhook, schedule, CLI, API). \\
\fld{finding\_type}          & Category (e.g.\ triage, IoC, control gap). \\
\fld{severity}              & Severity class. \\
\fld{confidence}            & Calibrated confidence score. \\
\fld{status}                & Lifecycle state (open, acknowledged, accepted, rejected, corrected). \\
\fld{summary}               & Short human-readable summary. \\
\fld{detailed\_analysis}     & Structured analysis body. \\
\fld{evidence}              & References (not copies) to supporting evidence. \\
\fld{recommended\_actions}   & Proposed actions, each with reversibility class. \\
\fld{hitl\_requirement}      & Required HITL tier (Section~\ref{sec:hitl}). \\
\fld{related\_findings}      & Links to upstream/related findings. \\
\fld{regulatory\_mappings}   & DORA, NIS2, GDPR, PSD2, 4AML, AI Act, ISO 27001, NIST CSF, PCI-DSS controls. \\
\fld{tool\_calls}            & Adapter calls and arguments used. \\
\fld{retrieval\_refs}        & RAG and graph queries used. \\
\fld{model\_refs}            & Model identifier and version. \\
\fld{metadata}              & Extension fields. \\
\bottomrule
\end{tabular}
\end{table}

The runtime stores \emph{references} to sensitive evidence rather than uncontrolled copies wherever feasible. This keeps the finding portable across the dashboard, the Reporting subagent, and the audit log without amplifying exposure of regulated data.

\subsubsection*{Worked example: suspicious-email triage}
Listing~\ref{lst:finding-example} shows the attribution-bearing fields of a finding produced by the Threat Intelligence subagent for a suspicious-email triage workflow (other Table~\ref{tab:finding-schema} fields elided).

\begin{lstlisting}[
  float,
  language={},        % or 'yaml', 'json', etc., if you like
  caption={Attribution-bearing fields of a populated finding. Sensitive content is held by reference (e.g.\ \texttt{evt:mailstore/msg/...}) rather than copied.},
  label={lst:finding-example}
]
summary: "Likely impersonation of client X
          requesting EUR 184k transfer."
severity: "medium"  confidence: 0.71
evidence: [evt:mailstore/msg/.../4781,
           evt:siem/alert/EML-7726,
           evt:client-baseline/X/lang-profile]
retrieval_refs: [rag:org-profile#client-X,
                 rag:phishing-playbook#impersonation,
                 graph:attck/T1566.002]
tool_calls: [siem.query{alert_id:EML-7726},
             mailstore.fetch_headers{msg_id:4781},
             cti.lookup_sender_domain{...}]
regulatory_mappings: [PSD2:Art97-SCA,
                      DORA:Art17-incident-reporting,
                      GDPR:Art32]
hitl_requirement: "tier-2"
recommended_actions: [draft-DORA-notification,
                      hold-payment]   (both reversible)
\end{lstlisting}

Every claim in \texttt{summary} and \texttt{detailed\_analysis} re-grounds through \texttt{evidence} and \texttt{retrieval\_refs} (supporting the unsupported-claim metric of Table~\ref{tab:eval}); \texttt{regulatory\_mappings} mechanically feeds the Reporting subagent's DORA Article~17 draft; and medium severity at 0.71 confidence triggers the Tier~2 routing of Table~\ref{tab:hitl}, halting the workflow for analyst countersignature.

\subsection{HITL tiers}\label{sec:hitl}
Decisions over agent actions follow three tiers. Routing is determined by the Main Agent on the basis of action class, finding severity, finding confidence, and reversibility, and is recorded in the audit log together with the routing rationale. Table~\ref{tab:hitl} maps action classes to tiers and to the artefact required to proceed.

\begin{table}[t]
\centering
\caption{HITL tier mapping. Tier assignment is recorded as part of the finding and audited.}
\label{tab:hitl}
\small
\begin{tabular}{p{0.10\columnwidth} p{0.46\columnwidth} p{0.32\columnwidth}}
\toprule
\textbf{Tier} & \textbf{Action class} & \textbf{Required artefact} \\
\midrule
Tier 1 & Reversible enrichment, triage, summarisation, IoC lookup, scoped retrieval. & None (autonomous). \\
Tier 2 & Confidence-gated actions; medium impact; reversible containment; draft regulatory notification. & Analyst countersignature. \\
Tier 3 & High-severity, low-reversibility, customer-impacting, regulatory-binding, or containment-with-side-effects actions. & Explicit, named approval; recorded justification. \\
\bottomrule
\end{tabular}
\end{table}

Tier assignments map directly to the EU AI Act Article~14 requirement for human oversight of automated decision support and to the DORA expectation of an audit trail covering operational decisions taken under regulatory deadlines.

\subsection{Append-only audit}
The audit log records every prompt, retrieval, graph query, tool call, model call, finding, HITL decision, report, and access denial. Each event carries the Security Context reference under which it occurred. The log is append-only; corrections are recorded as additional events that reference the original event identifier rather than as in-place edits.

\section{Extension Model and Evaluation Plan}\label{sec:extension}

\subsection{Pack types}
The architecture grows by adding packs (plugins) rather than by changing the Runtime Core. \emph{Connector packs} integrate external systems (SIEM/XDR, email, filesystem, Git/CI, ticketing, CTI feeds, vulnerability scanners, databases, optional MCP servers). \emph{Subagent packs} add specialist roles (Threat Intelligence, Compliance Verification, Incident Response, Reporting, Vulnerability Assessment, Behavioural Analysis, Forensic Analysis, DevSecOps and Code Analysis). \emph{Skill packs} add narrow workflows such as phishing triage, DORA incident notification, AML transaction graph analysis, PCI-DSS code review, forensic timeline reconstruction, or eBPF anomaly summarisation. \emph{Context packs} add regulatory corpora, ATT\&CK/CVE/CWE/CAPEC graphs, internal policies, playbooks, prior incidents, and asset inventories.

\subsection{Pack governance and lifecycle}\label{sec:pack-governance}
Each pack carries a signed manifest that declares its name, semantic version, authorship, public key, and the runtime claims it requires: tools, RAG namespaces, graph scopes, model classes, and skill dependencies. Verification keys are held in an operator-managed registry; an unsigned pack or a pack whose manifest claims exceed the operator's registered scope is refused at install time. A pack with new tool or scope claims can be \emph{installed} but cannot be \emph{enabled} without an explicit operator approval recorded in the audit log.

The pack lifecycle emits five audit events: \texttt{pack.install}, \texttt{pack.verify} (passed or failed), \texttt{pack.enable}, \texttt{pack.disable}, and \texttt{pack.upgrade}. Each event carries the manifest hash, the operator identity, and the Security Context under which the action was performed. Skill and connector packs additionally record evaluation hashes, allowing post-hoc correlation between a finding and the pack version that produced it. This structure addresses the supply-chain risks identified in the agent-security literature~\cite{ferrag2025prompt,ref-survey-llm-sec} without constraining the rate at which capabilities can be added.

\subsection{Evaluation Plan}\label{sec:evaluation}

The paper does not report production validation. Instead, we specify a falsifiable evaluation protocol applicable to the implementable slice (Section~\ref{sec:slice}). Each criterion is stated as a question, paired with a measurable metric, and given a pass criterion that determines whether the architecture's claim holds. Table~\ref{tab:eval} summarises the criteria across four dimensions.

\begin{table*}[ht]
\centering
\caption{Falsifiable evaluation plan with metric-level pass criteria.}
\label{tab:eval}
\small
\newcommand{\fld}[1]{\texttt{\seqsplit{#1}}}
\begin{tabular}{p{0.14\textwidth} p{0.30\textwidth} p{0.27\textwidth} p{0.23\textwidth}}
\toprule
\textbf{Dimension} & \textbf{Question} & \textbf{Metric} & \textbf{Pass criterion} \\
\midrule
Architecture readiness   & Can a new organization be onboarded without code changes?
                         & Lines-of-code changed; manual-config touchpoints
                         & 0 LOC; all changes confined to runtime config and organization profile. \\

Architecture readiness   & Can the same workflow run from every entry point (UI, CLI, alert, webhook, scheduler)?
                         & Workflow completion rate across entry points, per workflow
                         & $\geq$ 95\% completion on canonical workflows, no entry-point-specific code paths. \\

Architecture readiness   & Can the model provider be swapped without subagent changes?
                         & Code diff under provider swap (one local, one private API)
                         & Diff confined to the Model Gateway; subagent definitions unchanged. \\

Architecture readiness   & Can packs be enabled independently?
                         & Service restart count to enable a pack
                         & 0 restarts; install/enable recorded as audit events. \\

Security \& governance   & Is the Security Context enforced at every documented enforcement point (Table~\ref{tab:enforcement})?
                         & Pass rate on a denial test suite (one test per enforcement point $\times$ scope-violation case)
                         & 100\% denial; 100\% audit-event coverage with violated invariant recorded. \\

Security \& governance   & Are tool calls deterministic and audited?
                         & Schema-validation rejection rate on malformed arguments; audit-trace coverage
                         & 100\% rejection of out-of-schema arguments; every executed call audited. \\

Security \& governance   & Is HITL routing correct?
                         & Macro-F1 against labelled action-class$\rightarrow$tier mappings
                         & $\geq$ 0.9 macro-F1 on the canned scenario set. \\

Output quality           & Are findings fully attributed?
                         & \% findings with non-empty \fld{evidence}, \fld{retrieval\_refs}, \fld{tool\_calls}, \fld{model\_refs}, \fld{confidence}
                         & $\geq$ 95\%. \\

Output quality           & Are reports free of unsupported claims?
                         & \% report claims linked to at least one finding/evidence reference (LLM-judged + spot-checked)
                         & $\geq$ 90\% supported. \\

Output quality           & Do analyst corrections reach scoped memory without contaminating raw evidence?
                         & Memory-write isolation tests
                         & 100\% writes confined to scoped memory namespaces. \\

Operational              & Mean time to diagnose a failed workflow given the audit trail.
                         & Wall-clock time-to-root-cause on benchmark failures
                         & $<$ 5 minutes on benchmark scenarios. \\

Operational              & Connector health reporting completeness.
                         & \% connectors emitting health events on a fixed cadence
                         & 100\%. \\

Operational              & Scheduler backlog and latency stability.
                         & p95 routine-launch latency under nominal load
                         & $<$ 30 seconds. \\
\bottomrule
\end{tabular}
\end{table*}

We instantiate the implementable slice on a single workstation, ingest a fixed synthetic corpus (anonymised DORA-shaped incident emails, sample SIEM alerts, and an anonymised CVE list), and run six benchmark tasks: suspicious email or order triage, DORA incident summary drafting, threat-intelligence enrichment over an IoC, compliance mapping for a sample finding, daily posture summarisation, and a denied-access test. Each metric in Table~\ref{tab:eval} is reported under at least three model backbones (one local model, one private API endpoint, one stub returning canned outputs) to verify the model-agnostic requirement. The denied-access test is repeated for every row of Table~\ref{tab:enforcement}. Results, audit-log excerpts, and pack manifests of the configuration under test will be released alongside any future systems-paper submission.

\section{Discussion and Limitations}\label{sec:discussion}

The main trade-off is choosing one shared runtime rather than independent agent services. A shared runtime improves policy consistency, auditability, state management, and implementation simplicity. Independent services may become useful for scaling or isolation later, but treating them as the default would complicate the initial system without improving the core research claim.

The MCP-as-peer-adapter choice (Section~\ref{sec:architecture}) is a deployment hierarchy rather than an architectural one: native and CLI adapters are operationally preferred for high-audit paths, MCP connectors for dynamic-discovery paths, but all three traverse the same enforcement and audit surface.

We acknowledge two principal limitations. First, the contribution is architectural: the runtime contract, the Security Context, the Finding Schema, the HITL tier mapping, and the pack-governance lifecycle are specified but not yet empirically validated under operational load. The evaluation plan in Section~\ref{sec:evaluation} is falsifiable but has not been executed. As future work, we will instantiate the implementable slice (Section~\ref{sec:slice}) and report results against the metric-level pass criteria of Table~\ref{tab:eval}.

Second, several capabilities are deliberately treated as extension paths rather than runtime assumptions. Integrating each as connector or context packs while preserving the Security Context contract and the audit-event coverage of Tables~\ref{tab:enforcement} and~\ref{tab:hitl} is left to future work; multi-tenant operation requires the \texttt{tenant\_id} extension sketched in Section~\ref{sec:org-model}.

\section{Conclusion}\label{sec:conclusion}

Regulated cybersecurity operations need more than isolated LLM-agent capabilities. They need an organization-scoped runtime that can govern context, tools, state, memory, findings, reports, and audit, and that can keep that context current under the same governance contract. This paper proposed a local-first, model-agnostic architecture with one shared Runtime Core, logical specialist subagents, Security Context enforcement, governed adapters, structured findings, HITL tiers, extension packs, and an event-driven update loop that treats event-time and query-time access as duals of one contract. The design is intentionally implementation-aware but does not require production validation to be useful as an architecture proposal. The next step is to instantiate the implementable slice and evaluate architecture readiness, policy enforcement, evidence traceability, output quality, and operational observability.

\section*{Acknowledgment}
This work has received funding from the European Union under the DIGITAL EUROPE Programme, grant agreement No.~101249596 (CyberAId).\\
The authors used the AI tool Claude Opus 4.7 for grammar and style editing of the manuscript and Gemini NanoBanana to assist in figure generation, with all content reviewed and verified by the authors.

\bibliographystyle{IEEEtran}
\bibliography{references}

\end{document}